\documentclass[12pt]{article}

\usepackage[english]{babel}
\usepackage{amssymb}
\usepackage{latexsym}
\usepackage[round]{natbib}
\usepackage{url}
\usepackage{color}
\usepackage[font=small,format=hang]{caption}
\usepackage[final]{graphicx}
\usepackage{xr}

\newcommand{\prob}{\mathrm{Pr}}
\newcommand{\vat}{\mathbb{E}}

\newcommand{\bc}{\begin{center}}
\newcommand{\ec}{\end{center}}
\newcommand{\bit}{\begin{itemize}}
\newcommand{\eit}{\end{itemize}}
\newcommand{\be}{\begin{eqnarray*}}
\newcommand{\ee}{\end{eqnarray*}}
\newcommand{\ben}{\begin{eqnarray}}
\newcommand{\een}{\end{eqnarray}}

\newcommand{\g}{\,\vert\,}

\usepackage{color}
\renewcommand{\baselinestretch}{2}

{

	\usepackage[
top    = 1in,
bottom = 1.5in,
left   = 1in,
right  = 1in]
{geometry}

\newtheorem{lemma}{Lemma}[section]

\begin{document}
\title{Objective Bayesian Model Discrimination \\ in Follow-up Experimental Designs}
\author{GUIDO CONSONNI \and LAURA DELDOSSI\\
Dipartimento di Scienze Statistiche\\
Universit\`{a} Cattolica del Sacro Cuore\\
Largo Gemelli, 1; 20123 Milan, Italy\\
\url{guido.consonni@unicatt.it; laura.deldossi@unicatt.it}
}
\date{  }
\maketitle

\begin{abstract}
   An initial screening experiment may lead to ambiguous conclusions regarding the factors which are active in explaining the variation of an outcome variable: thus adding follow-up runs  becomes necessary.
We propose a  fully Bayes objective approach to follow-up designs, using prior distributions suitably tailored to model selection.
 We adopt a  model criterion based on  a weighted average of Kullback-Leibler divergences between   predictive distributions for all possible pairs of models.
When applied to real data, our method produces results which compare favorably to previous analyses  based on subjective weakly informative priors. Supplementary materials are available online.

   \vspace{0.5cm}
\end{abstract}

\section{Introduction}

%

In screening designs the objective is to discover which of the many potential factors are really active, i.e. contribute to explain the  variability of a response variable.


 In this context, it is customary to assume that the response  follows a normal linear  regression model, where the  predictors are
the  model-specific main effects together with all interactions up to a specified order (usually two).
In this way, for each given set of active factors, there is associated one and only one linear model.
If one considers $k$  factors, there exist
$2^k$ distinct models, including  the null model (no  factor is active), and  the full  model (all factors are active).

%




We adopt a
 Bayesian approach, wherein each uncertain quantity (such as  model, parameter or future observation) is assigned a prior (distribution) which, in the light of data, is updated to a posterior. In particular the Bayesian approach produces a full posterior distribution on the space of all models,  unlike in  frequentist model selection procedures (e.g. AIC, BIC, or penalized regression methods such as the Lasso).

Often screening designs are based on a limited number of runs,   and they may not lead to unequivocal conclusions as to which factors are active,
  because the posterior probability on model space is not sufficiently concentrated on a few models; and similarly for the induced posterior probability that each factor is active.
  As a consequence, extra runs are needed to resolve this ambiguity. The issue then becomes finding the   combination of factor levels which best discriminates among rival models, and hence factors. This brings us to optimal follow-up designs, which is the core of this paper.
  In this context, the following intuition can be helpful:
 a new experiment is most useful whenever the predicted response varies widely across  models, because this feature will facilitate model comparison. Accordingly,  the follow-up runs are chosen so as to maximize a \emph{model  discrimination} (MD)
criterion, see
\citet{Meye:Stei:Box:1996}.

To compute the posterior probability on each model, one requires a prior on model space, as well as a parameter prior on the  space of parameters (conditionally on each single model).  
A notorious difficulty associated with Bayesian  model determination is its  sensitivity to parameter priors; see \citet[ch. 7]{Ohag:Fors:2004}.   This remark, and the practical difficulty of specifying distinct subjective priors for each of the entertained models,   suggest to adopt an \emph{objective} Bayes approach \citep{Berg:Peri:2001}.
The latter program however cannot be carried out using
 standard
 noninformative priors for estimation purposes (if for no better reason that  they are typically improper); on the other hand
proper weakly informative priors, as  implemented  for instance in
   \citet{Meye:Stei:Box:1996}, are   also   questionable  for Bayesian model choice (high  sensitivity to prior specification of tuning parameters being an issue); for further discussion see \cite{Peri:2005}.

%


In this paper we address the problem of choosing follow-up experiments for optimal discrimination among factorial models,
using  a fully  objective Bayesian approach.
This seems particularly attractive at the screening stage, especially if  prior information is weak.
Specifically,
we seek to maximize an MD criterion which is a weighted combination of Kullback-Leibler (KL) divergences  between   predictive distributions (for future follow-up observations conditionally on the available data) for all pairs of models, where the weights are the posterior probabilities of the corresponding model pair; see \cite{Box:Hill:1967} for a derivation of this criterion using the notion of expected change
in entropy between input and output.
A related approach  was used by \citet{Bing:Chip:inco:2007} for identifying most promising screening designs. Their MD criterion however uses the  Hellinger distance, rather than KL-divergence, between pairs of (prior) predictive distributions.
Because of the structure of MD we decouple the problem into two separate sub-problems: i) finding the posterior probability on model space (which provides  the weights of the MD); ii) finding the predictive distributions of future observations (required to compute the KL divergences).


The  rest of this paper is organized as follows:
 Section \ref{sec:objective priors}  presents the  objective model choice priors;
Section \ref{sec:model discimination}  introduces the model discrimination criterion; Section \ref{sec:applications} applies the methodology to a variety of data sets, and provides comparison with previous analyses.
Finally Section \ref{sec:discussion} contains a brief discussion.

   \section{Objective model choice priors}

  \label{sec:objective priors}

\subsection{Model assumptions}

    Consider $k$  categorical factors,
    and
    $n$ experimental runs for specific  combinations of the factor levels.
    Let $M_i$ be a model which specifies a set of  $f_i$  active factors ($0 \leq f_i \leq k$) for the response $y$ 
\begin{eqnarray}
\label{linearmodelMi}
      y \g \beta_0, \beta_i, \sigma, M_i \sim N_n(X_0\beta_0 +X_i \beta_i, \sigma^2 I_n),
    \end{eqnarray}
      where $X_0$ represents an $n \times t_0$ design matrix containing  variables  which appear in all models. Typically,
    $X_0=1_n$, the $n$-dimensional unit vector; occasionally however, we may want to consider more general versions for $X_0$.
      Both $\beta_0$ and  $\sigma^2$ are regarded as parameters which are \lq \lq common\rq \rq{} across models, while $\beta_i$ is the model-specific vector of regression parameters.
    The $t_i$ columns of the matrix $X_i$ contain  suitable terms representing main effects and  interactions of selected factors.
    The model matrix $[X_0 \vdots X_i]$ is assumed to be of full column rank, so that the number of linearly independent terms in the regression structure cannot exceed $n$.
    In fractional factorial designs this means that only estimable (non aliased) interactions, up to a desired order, are introduced besides the main effects, conditionally to the constraint that $n>t_0+t_i$.

%

\subsection{Prior distribution on model space}
\label{subsec: prior on model space}

In this subsection we consider  in greater detail  the
    prior on model space.
    A typical assumption is that each factor is active (i.e. its effect will be included in any particular model) with some probability $\pi$  independently of the other factors.
    If $M_i$ contains $f_i$  active factors, $f_i\in \{0, 1, \ldots, k \}$, then
   \begin{eqnarray}
   \label{probMigivenpi}
   \prob(M_i \g \pi)=\pi^{f_i}(1-\pi)^{k-f_i}.
   \end{eqnarray}
       Of course $\pi$ is unknown, and from a Bayesian perspective  it should be regarded as an uncertain quantity with its own distribution. Assuming that $\pi \sim Beta(a,b)$,  then integrating (\ref{probMigivenpi}) with respect to this prior yields
       \begin{eqnarray}
\label{probMigeneral}
      \prob(M_i)=\int_{0}^1  \pi^{f_i}(1-\pi)^{k-f_i} p(\pi) d\pi=B(a+f_i, b+k-f_i)/B(a,b),
   \end{eqnarray}
   where $B(\cdot, \cdot)$ is the usual beta function.
 Priors belonging to  family (\ref{probMigeneral}) incorporate a multiplicity adjustment component; see  \citet{Scot:Berg:2010} for an extensive study.


\subsection{Parameter priors}
\label{subsec: priors on parameter space}

Consider  the comparison of  two nested models (so that the sampling family under one model is a special case of the other) through the Bayes factor.
  If one starts with an objective prior developed for estimation purposes, such as  the Jeffreys or reference prior, a difficulty arises: since these priors are typically improper, the  Bayes factor is undefined.
  Several attempts have been made to circumvent this problem: intrinsic Bayes factor \citep{Berg:Peri:1996},  fractional Bayes factor \citep{Ohag:1995},  intrinsic priors   \citet{Case:More:2006}, expected posterior prior \citep{Pere:Berg:2002}; see  \citet{Peri:2005} for  a comprehensive review.

  Another approach, whose precursor was essentially \citet{Jeff:1961}, is to develop a \emph{proper} prior for the parameter \lq \lq specific\rq \rq{} to the current model
  based on some reasonable intuition, and then test its reasonableness in specific settings  using simulation studies and possibly theoretical results.  Examples, with reference to variable selection in normal linear models, include  \citet{Zell:Siow:1980} on $g$-priors, \cite{Lian:EtAl:2008}, \cite{Clyd:Ghos:Litt:2011}, \cite{Maru:Geor:2011}; see also \citet{Baya:Garc:2007} for generalized linear models.

  Very recently,  a  different  approach was proposed by \cite{Baya:Berg:Fort:Garc:2012}, where  \emph{criteria} that should be satisfied by any model choice prior are first laid out in  generality, with special consideration for the objective case. Next,  one seeks  priors which satisfy these requirements, in the specific setting under investigation. We find this research strategy convincing, and adopt it in this paper to propose a solution for the follow-up experiment.

Recall the structure of model $M_i$  presented in (\ref{linearmodelMi}).
On the other hand the null model $M_0$ prescribes $y \g \beta_0, \sigma \sim N_n(\beta_0 X_0, \sigma^2 I_n)$.
Consider the following
hierarchical $g$-prior for model choice
\begin{eqnarray}
\label{p^R}
p^R(\beta_0, \beta_i, \sigma \g M_i)=
p(\beta_0,  \sigma)p^R(\beta_i \g \beta_0, \sigma \g M_i)=
\sigma^{-1} \times \int_0^{\infty}
N_{t_i}(\beta_i \g 0, g \Sigma_i)p^R(g \g M_i)dg,
\end{eqnarray}
where $p(\beta_0,  \sigma)$ is the prior on the common parameters shared by all models,
$\Sigma_i=\sigma^2 (V_i^{\prime}V_i)^{-1}$, $V_i=(I_n-X_0(X_0^{\prime}X_0)^{-1}X_0^{\prime})X_i$ and
\begin{eqnarray*}
\label{p^Rg}
p^R(g \g M_i)=\frac{1}{2} \left[  \frac{1+n}{t_i+t_0}\right]^{1/2}(g+1)^{-3/2}1_{(\frac{1+n}{t_i+t_0}-1, \infty)}(g),
\end{eqnarray*}
with $1_A(t)=1$ if $t \in A$  and 0 otherwise.
Prior (\ref{p^R}) has been shown to satisfy all \emph{desiderata} from an objective model choice perspective;
 see \citet[Section 2]{Baya:Berg:Fort:Garc:2012} (the superscript \lq \lq R\rq \rq{} stands for \lq \lq robust prior \rq \rq{}).
Notice that (\ref{p^R}) is improper; however it scales appropriately  when compared to the null model $M_0$ so that the resulting Bayes factor for the comparison of $M_i$ against $M_0$ is meaningful. Its expression is
\begin{eqnarray}
\label{BF_i0}
\lefteqn{BF_{i0}(y)= \left[ \frac{n+1}{t_i+t_0} \right]^{-t_i/2}} \nonumber \\
&&
\hspace{-.5cm} \times \frac{Q_{i0}(y)^{-(n-t_0)/2}}{t_i+1}
{_2F_1}\left[ \frac{t_i+1}{2}; \frac{n-t_0}{2}; \frac{t_i+3}{2}; \frac{(1-Q_{i0}(y)^{-1})(t_i+t_0)}{n+1} \right]  ,
\end{eqnarray}
where $_2F_1$ is the standard hypergeometric function \citep{Abra:Steg:1964}, and $Q_{i0}(y)=SSE_i(y)/SSE_0(y)$
is the ratio of the sum of squared errors of models $M_i$ and $M_0$.




   \section{The model discrimination criterion}
   \label{sec:model discimination}

%
%

Assuming that one of the entertained models is true,
the posterior probability  of each model $M_i$ can be written in the convenient form
\begin{eqnarray}
\label{probMiGiveny}
\prob(M_i \g y)=\frac{BF_{i0}(y)P_{i0}}{1+\sum_{j \neq 0} BF_{j0}(y)P_{j0}},
\end{eqnarray}
where $P_{j0}$
is the prior odds of model $M_j$ relative to $M_0$ implied by
(\ref{probMigeneral}), and $BF_{j0}(y)$ is defined in (\ref{BF_i0}).
In particular we adopt (\ref{probMigeneral}) with $(a=1,b=1)$ for which $P_{j0}=f_j !(k-f_j)!/k!$.

Having computed $\prob(M_i \g y)$, $i=1, \ldots, 2^k$,  a useful   by-product  is the posterior probability $P_A(y)$ that  factor $A$ say is active; namely
 $
 \sum_{ \{M_j:\, \mbox{factor A is active} \}}\prob(M_j \g y)$.

In order  to determine the $n^*$ follow-up runs which best discriminate among potential explanatory models,
   \citet{Meye:Stei:Box:1996} suggested to maximize the following  model discrimination (MD) criterion
   \ben
   \label{MD-PredictiveCrit}
      MD=\sum_{ i \neq j} \prob(M_i|y)\prob(M_j|y)KL(m(\cdot|y, M_i), m(\cdot|y, M_j)),
   \een
   where  $m(\cdot|y, M_i)$ is the (posterior) predictive density for the  vector of follow-up observations, 
   and
   \begin{eqnarray}
   KL(f,g)= \int f(x) \log \frac{f(x)}{g(x)}dx
   \end{eqnarray}
   is the Kullback-Leibler divergence of the density $f$  from $g$.
    Notice that  MD is a weighted average of the  KL-divergences between all pairs of  predictive distributions for the follow-up observations.

Adopting a standard reference prior
$p^N(\beta_0, \beta_i, \sigma \g  M_i) \propto 1/\sigma$  for prediction purposes leads to a closed form expression for the MD criterion, which we label OMD (Objective MD). This is given by
\ben
\label{MDfinal}
&& OMD=   \\
&&   \sum_{ i \neq j} \prob(M_i|y)P(M_j|y)
\frac{1}{2} \left\{ tr(V_j^{* \,-1} V_i^*) + \frac{n-t_i-t_0}{SSE_i}(\hat{y}_i^*-\hat{y}_j^*)^{\prime} V_j^{* \,-1} (\hat{y}_i^*-\hat{y}_j^*)
  -n^* \right\}, \nonumber
\een
where $\prob(M_i|y)$ is defined in (\ref{probMiGiveny}),   $SSE_i$ is the usual residual sum of squares under model $M_i$,  and the remaining quantities are defined in (\ref{y_i*V_i*}) of Appendix A in the Supplementary Materials, which includes further details.
To find the best follow-up runs one has to maximize OMD over all possible $n^*$ combinations with repetition from the set of runs of the full factorial design.

\citet{Meye:Stei:Box:1996} evaluate MD using
 (\ref{probMigivenpi}) with  $\pi$  equal to a fixed small value  (the recommended choice was $\pi=0.25$) to induce factor sparsity in model selection.
 Additionally, they chose $p(\beta_0, \sigma) \propto 1/\sigma$ as we did, while adopting  a  proper weakly informative Gaussian prior on  $\beta_i\g \beta_0, \sigma, M_i$, wherein each component of $\beta_i$ is assigned a normal distribution with zero expectation and standard deviation  $\gamma \sigma$, with $\gamma$ a tuning parameter, which need be specified by the user. We label the resulting criterion CMD (Conventional MD).

OMD has the advantage, with respect to CMD, of being  fully Bayes,   objective,  and based on principled model selection priors. In particular, there is no need to tune hyperparameters, which makes it especially attractive  from a practitioner's point of view.  In fact, on the one hand it is well known that   model selection is typically highly sensitive to the choice of hyperparameters;  on the other hand this  prior information is usually hard to elicit in screening experiments.

We have developed Fortran and  R-code  to find the optimal follow-up runs under OMD. This code relies on existing Fortran and R-code to carry out computations under CMD; see \citet{Meye:1996} and \citet{Barr:2013}.

\section{Applications}
\label{sec:applications}
\subsection{Injection molding experiment}
\label{subsec:injecion}


We first consider the experiment on the percentage shrinkage in an injection molding process described in
\citet[p. 398]{Box:Hunt:Hunt:1978}  which contains eight factors labeled A through H. This experiment was
 also analyzed  in \citet{Meye:Stei:Box:1996}. The plan is a $2^{8-4}$ fractional factorial resolution IV design with generators I=ABDH=ACEH=BCFH=ABCG. 

A preliminary analysis based on normal probability plots, and confirmed by a Bayesian analysis (be it conventional or objective), leads to the conclusion that the potential active factors can be reduced to four, namely  A, C, E and H.
 Accordingly, we follow \citet{Box:Hunt:Hunt:1978} and collapse the $2^{8-4}$ fractional factorial design on the above factors, thus  obtaining a replicated $2^{4-1}$ design with defining relation I=ACEH.

The posterior probabilities  that each factor is active are reported  in Table \ref{tab:Injection.Post.Prob.Factors.Active}.
 They are essentially uniform for the case of three-factor interactions (3FI) both under the conventional and the objective approach; this result is  only partly modified  in the case of  2FI,  where  factor A appears unlikely to be active under the conventional approach.
%
It appears that additional runs are needed both to resolve the ambiguity regarding factor A, and to further investigate the role of the remaining factors.


\renewcommand{\baselinestretch}{1}
\begin{table}[tbp] \footnotesize
\centering
\caption{{\protect\small \textit{Injection molding experiment.
Posterior probabilities that factors are active for the $2^{4-1}$ design}}}
\label{tab:Injection.Post.Prob.Factors.Active}%
\begin{tabular}{*{6}{c}}
\hline
\hline
& \multicolumn{2}{c}{2FI} & & \multicolumn{2}{c}{3FI } \\
\cline{2-3}
\cline{5-6}
Factor & Conventional Approach & Objective Approach & & Conventional Approach & Objective Approach\\
\hline $A$ & $0.18$ & $0.68$ & & $0.76$ & $0.87$ \\
 $C$ & $1$ &$1$ & & $0.76$ & $0.88$ \\
 $E$ & $1$ & $1$ & & $0.76$ &  $0.87$\\
 $H$ & $0.91$ &  $0.95$ & & $0.76$ & $0.87$ \\
 \hline
\end{tabular}%
\end{table}


Table \ref{tab:Injection.Candidate.Runs} in the Supplementary materials reports the  full $2^{4}$ design  in the factors A, C, E, H, with the corresponding runs in the $2^{8-4}$ fractional design.
 Assuming that $n^*=4$ follow-up runs have to be chosen, the number of possible follow-up designs  (with replication) from the 16 candidate runs  of the $2^4$ factorial  design in the factors A, C, E, H is 3876. 

 The five best designs identified by the OMD criterion of formula (\ref{MDfinal}),  along with those corresponding to the CMD criterion are shown in Table \ref{tab:Injection.Top5},  separately for models having 2FI and 3FI. The CMD criterion was applied using the recommended  settings $\pi=0.25$ and $\gamma=2$, and without a block-effect to distinguish between screening and follow-up runs.
 We also report for completeness the  value of the criterion achieved by each single run. Please notice that these values are  meaningful for comparison purposes within each criterion, but not between criteria.

\renewcommand{\baselinestretch}{1}
\begin{table}[tbp] \footnotesize
\centering
\caption{{\protect\small \textit{Injection molding experiment.
Top five follow-up designs}}}
\label{tab:Injection.Top5}%
\begin{tabular}{*{10}{c}}
\hline
\hline
& \multicolumn{4}{c}{2FI} & & \multicolumn{4}{c}{3FI} \\
\cline{2-5}
\cline{7-10}
Model & CMD & runs  & OMD  & runs & & CMD & runs  & OMD  & runs \\
\hline
 $1$ & 11.23 & 9  12 13 16 & $51.12$ & 9 13 15 16 & & $88.37$ & 9 11 12 15 & $103.41$ & 9 10 11 13 \\
 $2$ & $11.08$ & 9  12 15 16  & $49.84$ &  9 12 13 16  & & $87.37$ &  9 12 12 15  & $100.39$ &  9 10 11 12  \\
 $3$ & $10.99$ & 11 12 15 16  & $49.53$ &  9  9 13 15  & & $87.35$ &  9 9 12 15  & $99.03$ &  9 10 10 11  \\
 $4$ & $10.92$ & 9  11 12 16  & $49.52$ &  9  9 13 16  & & $86.53$ &  9 12 14 15  & $98.34$ &  9 10 11 16  \\
 $5$ & $10.87$ & 12 13 15 16  & $48.22$ &  9  11 13 16  & & $83.87$ &  9 11 12 12  & $98.09$ &  9 10 11 11  \\
\hline
\end{tabular}%
\end{table}

       A common feature is that all follow-up runs belong to the set $\{9, 10, \ldots,16\}$, i.e.  the set of runs which were \textit{not} carried out in the initial screening experiment; this is reassuring because those runs were not able to discriminate sufficiently  among models.
 Some differences emerge   depending on the number  of FI allowed in the models as well as  the criterion  which is adopted (CMD and OMD), although runs 9 and 11 are broadly recurring.

\subsection{Reactor experiment}
\label{subsec:reactor}

In this subsection we consider the reactor experiment described in \citet[p. 376]{Box:Hunt:Hunt:1978}. Table \ref{tab:Reactor.Full.Fact.Design} in the Supplementary Materials reports the complete $2^{5}$  factorial design, including the value of the response variable. This feature makes this experiment especially attractive, because we can actually verify the effectiveness of our approach in identifying active factors, as we detail below.

\begin{figure}
\begin{center}
\includegraphics[width=6in]{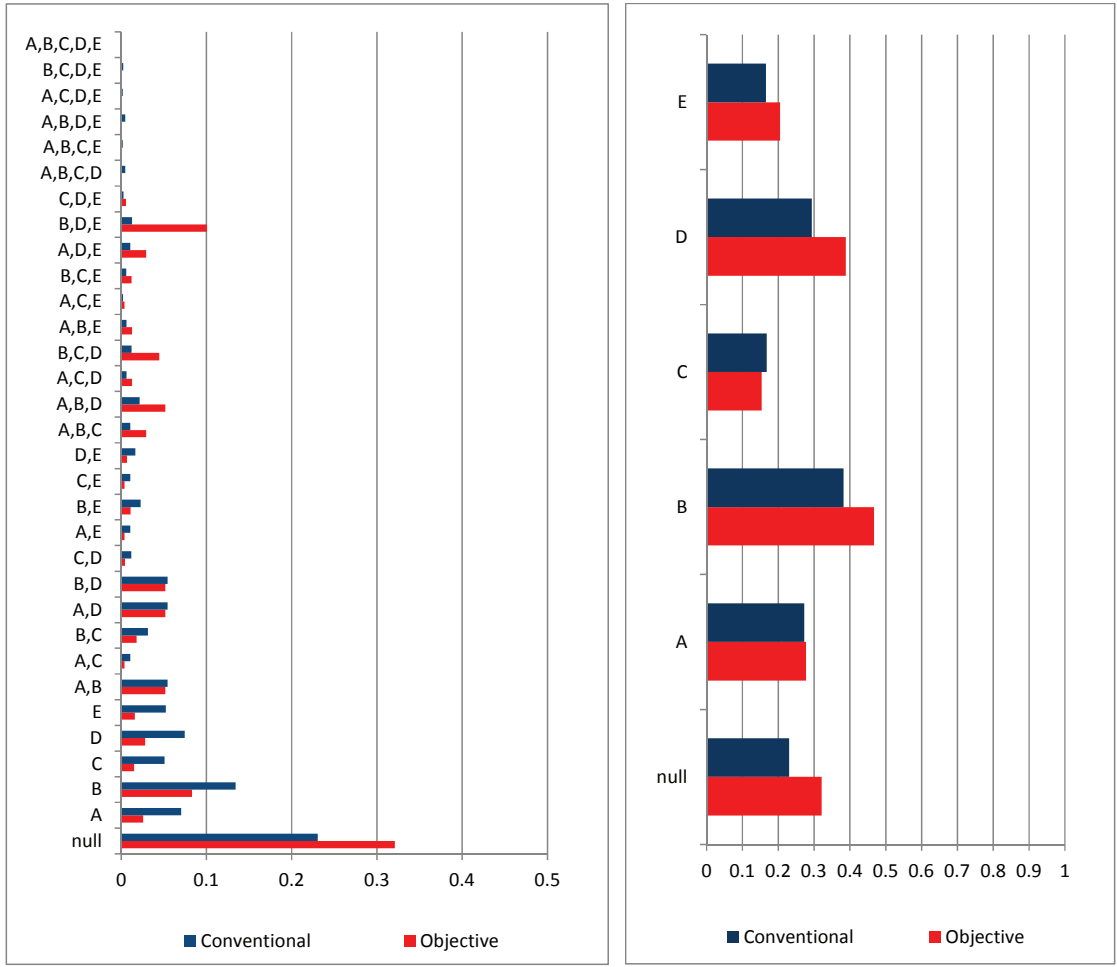}
\end{center}
\caption{{\itshape Reactor experiment. Posterior  probabilities of models and that factors are active (2FI)}}
\label{fig:Reactor.Post.Prob.Models.Factors.2FI}
\end{figure}

Following \citet[Section 3]{Meye:Stei:Box:1996},  we extract eight runs from the original experiment corresponding to the  $2^{5-2}$ Resolution III design with generators I=ABD=ACE,  and consider these runs as our initial screening design; see Table \ref{tab:Reactor.Screening.Design} in the Supplementary Materials.
The five highest posterior probability (top) models based on the objective Bayes approach are reported in Table \ref{tab:Reactor.Post.Prob.TopModelsAndActFac.2FI}
for models with 2FI. The corresponding results for the case of 3FI are reported in Table \ref{tab:Reactor.Post.Prob.TopModelsAndActFac.3FI} in the Supplementary Materials.
For the sake of comparison
we also included  the corresponding   results based on the conventional approach   derived in \citet{Meye:Stei:Box:1996} (setting $\gamma=0.4$  and $\pi=0.25$). The posterior probabilities of all models, and that the factors are active, are also displayed in Figure
\ref{fig:Reactor.Post.Prob.Models.Factors.2FI} for the case of 2FI (and Figure   \ref{fig:Reactor.Post.Prob.Models.Factors.3FI}
in the Supplementary Materials for the case of 3FI).

\renewcommand{\baselinestretch}{1}
\begin{table}[tbp] \footnotesize
\centering
\caption{{\protect\small \textit{Reactor experiment.
Posterior probabilities of top five  models and that factors are active (2FI)}}}
\label{tab:Reactor.Post.Prob.TopModelsAndActFac.2FI}%
\begin{tabular}{*{6}{c}}
\hline
\hline
 & \multicolumn{2}{c}{Conventional Approach} & & \multicolumn{2}{c}{Objective Approach} \\
\cline{2-3}
\cline{5-6}
 Model & Factors   &  Posterior probability & & Factors & Posterior probability  \\
\hline $1$ & $null$  & $0.23$ & & $null$ & $0.32$\\
 $2$ & $B$  & $0.13$ &  & $B,D,E$  & $0.10$ \\
 $3$ & $D$  & $0.07$ & & $B$  & $0.08$\\
 $4$ & $A$  & $0.07$ & & $A,D$  & $0.05$ \\
 $5$ & $A,D$  & $0.05$ & & $B,D$  & $0.05$ \\
\hline
\hline
 & \multicolumn{2}{c}{Conventional Approach} & & \multicolumn{2}{c}{Objective Approach} \\
\cline{2-3}
\cline{5-6}
 &  Factor & Posterior probability &  &  Factor & Posterior probability  \\
\cline{2-6}
 & $A$ & $0.27$  & & $A$ & $0.28$ \\
 & $B$ & $0.38$  & & $B$ & $0.47$ \\
 & $C$ & $0.17$   & & $C$ &  $0.15$\\
 & $D$ & $0.29$  & & $D$ & $0.39$  \\
 & $E$ & $0.17$   & &  $E$ & $0.21$  \\
 \hline
\end{tabular}%
\end{table}


 It appears from Figure \ref{fig:Reactor.Post.Prob.Models.Factors.2FI} that   the  objective Bayes prior tends to favor,  relative to the conventional approach,  the null model as well as a few models containing three factors. This is due to the different nature of the respective priors on model space.
 The posterior probabilities  that factors are active do not point to a clear-cut conclusion. The highest scoring factor (B)  does not even achieve the 50\% threshold; the remaining factors trail behind but each one has an appreciable probability of being active.  Extra runs are needed in order to solve what appears to be an ambiguous outcome.


To facilitate the comparison with \citet{Meye:Stei:Box:1996}, we chose to add $n^*=4$ follow-up runs.
For this problem there exist 52360  four-run designs  (with replications) from 32 candidates.
 The five best follow-up designs  selected by the  OMD, as well as the CMD, criterion  are shown in Table \ref{tab:Reactor.Top.Five.Follow-up.2FI} for the case of 2FI.
\renewcommand{\baselinestretch}{1}
\begin{table}[tbp] \footnotesize
\centering
\caption{{\protect\small \textit{Reactor experiment.
Top five follow-up designs (2FI)}}}
\label{tab:Reactor.Top.Five.Follow-up.2FI}%
\begin{tabular}{*{7}{c}}
\hline
\hline
Model & & CMD & runs  & & OMD  & runs \\
\cline{1-1}
\cline{3-4}
\cline{6-7}
 $1$ & & $0.5840$ & 4  10 12 26 & & $69.85$ & 11 15 26 29  \\
 $2$ & & $0.5821$ & 4  12 26 27 & & $69.73$ &  15 15 29 30    \\
 $3$ & &  $0.5800$ & 10 12 26 27 & & $69.71$ &  11  15 26 30   \\
 $4$ & & $0.5797$ & 4  11 12 26 & & $69.63$ &  11  15 29 30   \\
 $5$ & & $0.5792$ & 4 10 26 28  & & $69.42$ &  11  15 25 30   \\
\hline
\end{tabular}%
\end{table}
The best four runs under the OMD criterion only marginally overlap (run 26) with those obtained using CMD; on the other hand they do coincide when models with three-factor interactions are considered; see  Table \ref{tab:Reactor.Top.Five.Follow-up.3FI} in the Supplementary Materials.

\begin{figure}
\begin{center}
\includegraphics[width=6in]{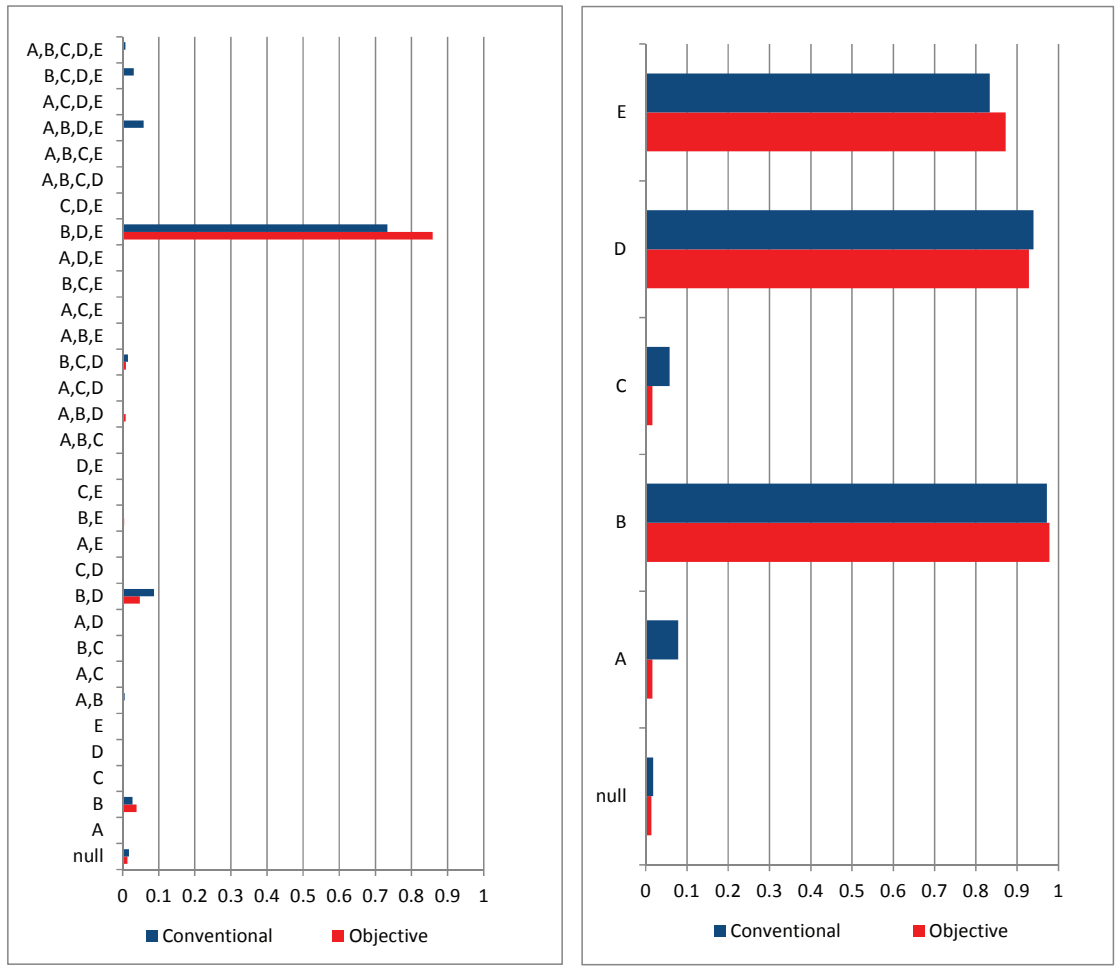}
\end{center}
\caption{{\itshape Reactor experiment. Posterior model probabilities of models and that factors are active based on  the combined screening and follow-up designs (2FI)}}
\label{fig:Reactor.Post.Prob.Combined.2FI}
\end{figure}

To validate the effectiveness of our approach, we re-run the analysis using all 12 runs (screening \textit{and} follow-up). To account for potential different experimental conditions,  a block effect was added in each linear model.
For models having 2FI, the results are summarized in Table \ref{tab:Reactor.Post.Prob.Combined.2FI}, and also  displayed in Figure \ref{fig:Reactor.Post.Prob.Combined.2FI}.

\renewcommand{\baselinestretch}{1}
\begin{table}[tbp] \footnotesize
\centering
\caption{{\protect\small \textit{Reactor experiment.
Posterior probabilities of top five models and that factors are active based on the combined screening and follow-up designs (2FI)}}}
\label{tab:Reactor.Post.Prob.Combined.2FI}%
\begin{tabular}{*{6}{c}}
\hline
\hline
 & \multicolumn{2}{c}{Conventional Approach} & & \multicolumn{2}{c}{Objective Approach} \\
\cline{2-3}
\cline{5-6}
 Model & Factors   &  Posterior probability & & Factors & Posterior probability  \\
\hline $1$ & $B,D,E$  & $0.73$ & & $B,D,E$ & $0.86$\\
 $2$ & $B,D$  & $0.09$ & & $B,D$  & $0.05$ \\
 $3$ & $A,B,D,E$  & $0.06$ & & $B$  & $0.04$\\
$4$ & $B,C,D,E$  & $0.03$ & & $null$  & $0.01$ \\
 $5$ & $B$  & $0.03$ & & $B,C,D$  & $0.01$ \\
\hline
\hline
 & \multicolumn{2}{c}{Conventional Approach} & & \multicolumn{2}{c}{Objective Approach} \\
\cline{2-3}
\cline{5-6}
 &  Factor & Posterior probability &  &  Factor & Posterior probability  \\
& $A$ & $0.08$ & & $A$ & $0.02$ \\
 & $B$ & $0.97$ & & $B$ & $0.98$ \\
 & $C$ & $0.06$ &  & $C$ &  $0.02$\\
 & $D$ & $0.94$  & & $D$ & $0.93$  \\
 & $E$ & $0.83$  &  & $E$ & $0.87$  \\
 \hline
\end{tabular}%
\end{table}

It now appears clearly that the only model worth of consideration is  the one involving factors B, D and E; these results are also spelled out in the posterior probabilities that factors are active.
Table \ref{tab:Reactor.Post.Prob.Combined.3FI} and Figure \ref{fig:Reactor.Post.Prob.Combined.3FI}.
 in the Supplementary Materials illustrate the analysis for models involving three-factor interactions with results broadly similar to those obtained under the 2FI case, the main difference being that factor E appears less likely to be active.
The above results obtained on the basis of 12 runs are in agreement with those that emerge from the normal probability of the contrasts based on the \textit{complete} set of 32 runs; see Figure \ref{fig: Reactor.Normal.Prob.Contrasts} in the Supplementary Materials.

Clearly the follow-up runs greatly contributed to differentiate among factors in terms of their likely activity. Which of the two approaches, conventional or objective, did a better job? Table \ref{tab:Reactor.Shannon} offers an answer. It computes the normalized  Shannon heterogeneity index on the posterior distribution of models after: (1) the screening experiment,   and (2) the combined screening and follow-up experiment.
 Clearly the index is lower in the latter situation, reflecting a reduced heterogeneity (increased concentration). We can see that our objective criterion not only scores lower after (1) and (2) than the conventional one, but it also produces a greater relative reduction  (71\% against 59\%).

\renewcommand{\baselinestretch}{1}
\begin{table}[tbp] \footnotesize
\centering
\caption{{\protect\small \textit{Reactor experiment.
Shannon hetereogenity of model posterior probabilities}}}
\label{tab:Reactor.Shannon}%
\begin{tabular}{{lcccc}}
\hline
\hline
& & Conventional Approach & & Objective Approach  \\
\cline{1-1}
\cline{3-3}
\cline{5-5}
(1): Screening experiment & & $0.79$ & & $0.74$ \\
(2): Screening and follow-up experiment & & $0.32$ & & $0.21$     \\
\\
Relative reduction between (1) and (2) & &  $59\%$ & &  $71\%$  \\
\hline
\end{tabular}
\end{table}

 A similar exercise  was performed  with respect to the  posterior probabilities that the factors are active. In this case, one can no longer use Shannon heterogeneity because the probabilities do not sum to one (the events are not incompatible). Accordingly, we chose the coefficient of variation. In this case situation (2) corresponds to a greater variation. Again OMD provides a higher score than CMD both in case  (1) and (2), even though CMD provides a greater improvement in relative terms; see Table \ref{tab:Reactor.Coefficient.Of.Variation}.

\renewcommand{\baselinestretch}{1}
\begin{table}[tbp] \footnotesize
\centering
\caption{{\protect\small \textit{Reactor experiment.
Coefficient of variation of posterior probabilities that factors are active}}}

\label{tab:Reactor.Coefficient.Of.Variation}%
\begin{tabular}{{lcccc}}
\hline
\hline
& & Conventional Approach & & Objective Approach  \\
\cline{1-1}
\cline{3-3}
\cline{5-5}
(1): Screening experiment & & $0.32$ & & $0.39$ \\
(2): Screening and follow-up experiment & & $0.72$ & & $0.80$     \\
\\
Relative increase between (1) and (2) & & $125\%$  & & $105\%$  \\
\hline
\end{tabular}%
\end{table}

\section{Discussion}
\label{sec:discussion}

In this paper we have developed an objective Bayesian method to obtain follow-up designs which are optimal in terms of predictive model discrimination.
In order to determine the posterior probability of  models,  we have employed  a multiplicity correction  prior on model space,  and a  principled model selection hierarchical-$g$ prior  on the parameters.
With regard to prediction, we have relied on a standard reference prior, which produces a closed-form expression for the model discrimination criterion, thus   greatly enhancing the computational speed of searching through the space of potential designs.
Employing different priors for model selection and prediction  implies
       that our model discrimination criterion will no longer   enjoy the  theoretical  properties described in the original contribution of  \citet{Box:Hill:1967}. However, it will do so at least approximately, because predictions based on the standard reference prior are themselves an approximation  to those computed using  the model selection prior; see  Appendix B of the Supplementary materials.
Finally, we remark that the practice of using distinct prior distributions for design and estimation-prediction dates back at least to \citet{Tsut:1972}. For a more recent example see \citet{Han:Chal:2004}, and references therein, where  the motivation is that distinct researchers, with different priors, may be involved in the design and estimation stage.


Our objective Bayes approach requires that the design matrix be of full rank.
This is in contrast to what happens in subjective Bayes approaches where this condition can be relaxed at the expense of having to specify a prior covariance matrix on the regression coefficients. Substantive prior information of this kind is usually unavailable, and conventional choices are problematic because model selection is highly sensitive to such prior inputs; see \citet{Berg:Peri:2001}.
The requirement that the design matrix be of full rank  implies that the set of models that can be entertained -for a given order of interactions-  may be smaller  than that of all potential models. This difficulty however can be typically overcome by omitting  models containing  higher-order interactions, or context variables (such as blocking). Since the main goal is obtaining the posterior probability of the active factors -rather than the posterior probability of the models- this simplification seems reasonable.

With regard to the prior on  model space presented in Subsection \ref{subsec: prior on model space},
 we adopted the values $(a=1,b=1)$. Recently the alternative  choice $(a=1, b=k+1)$ has been advocated to achieve a stronger  sparse modeling effect. This prior,   besides performing multiplicity adjustment, is also optimal in terms of concentration of the posterior distribution around the true model; see  \cite{Cast:Vaar:2012}.
  Having experimented with such prior,  the main difference is that the choice $(a=1$, $b=k+1)$ gives more weight to more parsimonious models, 
  relative to $(a=1, b=1)$; however, optimal follow-up runs, are broadly similar in the two cases.

The prior on model space adopted  in this paper relies on the assumption of effect forcing whereby if a set of factors is inserted in the model, then all interactions (up to the desired order) must be included.
%
One could relax the assumption of effect forcing, and consider a more flexible approach,  as   advocated in \citet{Bing:Chip:inco:2007}, through the incorporation of  prior opinions on structural aspects of effects such as  Effect sparsity, Effect hierarchy and  Effect heredity; see also \citet{Wolt:Bing:2011}.


The model discrimination criterion used in this work is based on the Kullback-Leibler divergence. Alternative divergence measures could be employed.    For instance, within the context of screening experiments,  \citet{Bing:Chip:inco:2007} suggest to use the  Hellinger distance, which is symmetric and bounded above. Symmetry is useful from the computational perspective,  because it avoids to sum over all pairs of distinct models, while a bounded index  makes calibration and interpretation easier. We could  implement our method using the Hellinger distance because its expression is also available in closed-form. The choice of the KL-divergence was mostly motivated for comparison purposes with results in the current literature.

\section*{ACKNOWLEDGMENTS}
The R-code  to find the optimal follow-up runs was developed by Marta Nai Ruscone, Dipartimento di Scienze Statistiche, Universit\`{a} Cattolica del Sacro Cuore, Milan.

We are indebted to the participants to the O-Bayes 2013 conference (December 15-17, 2013; Duke University) for useful comments on a preliminary version of this paper.
In particular we thank Veronika  R\v{o}ckov\'{a}  for a detailed discussion of our work, including priors on model space and the derivation of the model discrimination criterion, as well as Gonzalo Garc\'{i}a-Donato for pointing out the relationship between the posterior under the hierarchical $g$-prior and that based on the reference prior.

\section*{Supplementary Materials}
\begin{description}

\item[Appendix A:]
Derivation of KL-divergence between the predictive distributions for the follow-up runs under two models.

\item[Appendix B:] Relationship between the posterior distributions under the  hierarchical $g$-prior and the reference prior.

\item[Tables and Figures:] A collection of Tables and Figures complementing those in the main text.


\end{description}


\newpage
\subsection*{Appendix A: derivation of KL-divergence between the predictive distributions for the follow-up runs under two models}

Let $y^*$ denote the vector of observations for the $n^*$ follow-up runs. Under model $M_i$,   let
$\gamma_i^{\prime}=(\beta_0^{\prime}, \beta_i^{\prime})$,
and denote with $p^N(\gamma_i, \sigma^2 \g M_i)$ an objective estimation prior, where the superscript \lq \lq N\rq \rq{} stands for \lq \lq noninformative\rq \rq{}. Then
\be
m(y^* \g y, M_i)= \int \int f(y^* \g \gamma_i, \sigma^2, M_i)p^N(\gamma_i, \sigma^2 \g y, M_i) d\gamma_i d\sigma^2,
\ee
where $f(y^* \g \gamma_i, \sigma^2, M_i)=N_{n^*}(y^* \g Z_i \gamma_i, \sigma^2 I_{n^*})$ is the
usual Gaussian regression model having set
  $Z_i=[X_0 \vdots X_i]$.
Standard computations yield
\ben
\label{p^NGiveny}
&& p^N(\gamma_i, \sigma^2 \g y,M_i)
=p^N(\gamma_i \g \sigma^2, y, M_i) p^N( \sigma^2 \g y,M_i) \nonumber \\
&=& N_{t_i+t_0}(\gamma_i \g \hat{\gamma_i}, \sigma^2 (Z_i^{\prime}Z_i)^{-1})
IGa(\sigma^2 \g \frac{n-t_i-t_0}{2}, \frac{SSE_i}{2}),
\een
where $\hat{\gamma}_i$ is the OLS estimate of $\gamma_i$ and $IGa(t \g a,b)$ is the inverse gamma density having kernel $(1/t)^{a+1}\exp(-b/t)$.

As a consequence the predictive distribution of $y^*$, conditionally on $\sigma^2$ and under model $M_i$, can be written as
\ben
\label{predictivey*}
m(y^* \g \sigma^2, y, M_i)=N_{n^*}(y^* \g \hat{y}_i^*, \sigma^2 V_i^*),
\een
where
\ben
\label{y_i*V_i*}
\hat{y}_{i}^*=
Z_{i}^* \hat{\gamma}_i, \quad  V_i^*=I_{n^*}+Z_i^*(Z_i^{\prime}Z_i)^{-1}Z_i^{*\,^{\prime}}.
\een

To compute the KL divergences between pairs of predictive distributions appearing in formula (\ref{MD-PredictiveCrit}) of the paper, we proceed in two steps. First we evaluate the KL divergence conditionally on $\sigma$, and then we take the expectations with respect to the posterior distribution of $\sigma^2$.

Conditionally on $\sigma$, the predictive distributions are multivariate normal, and the following Lemma is useful.
\begin{lemma}
Let $m_0(\cdot)$ and $m_1(\cdot)$ be two $s$-dimensional multivariate Gaussian distributions with expectations $\mu_0$ and $\mu_1$ and covariance matrices $\Sigma_0$ and $\Sigma_1$. Then
\ben
\label{KLGaussian}
KL(m_0(\cdot), m_1(\cdot))=\frac{1}{2} \left\{ tr(\Sigma_1^{-1} \Sigma_0) + (\mu_1-\mu_0)^{\prime} \Sigma_1^{-1} (\mu_1- \mu_0) +\log \left( \frac{|\Sigma_1|}{|\Sigma_0|}\right) -s \right\}.
\een
\end{lemma}
As a corollary we  get
\ben
\label{KLPredictivesGivenSigma}
\hspace{-1.5cm} && KL(m(\cdot \g \sigma^2, y, M_i), m(\cdot \g \sigma^2, y, M_j))= \nonumber  \\
\hspace{-1.5cm} && \frac{1}{2} \left\{ tr(V_j^{* \,-1} V_i^*) + \frac{1}{\sigma^2}(\hat{y}_i^*-\hat{y}_j^*)^{\prime} V_j^{* \,-1} (\hat{y}_i^*-\hat{y}_j^*)
 + \log \left( \frac{|V_j^*|}{|V_i^*|}\right) -n^* \right\}.
\een

The last step involves an expectation with respect to the posterior distribution of $\sigma^2$. Since   $ \sigma^2 \sim IGa( \frac{n-t_i-t_0}{2}, \frac{SSE_i}{2})$, we  get
$\vat(1/\sigma^2 \g y, M_i)=(n-t_i-t_0)/SSE_i$.
Therefore
\ben
\label{KLPredictives}
\hspace{-1.5cm} && KL(m(\cdot \g  y, M_i), m(\cdot \g  y, M_j))= \nonumber  \\
\hspace{-1.5cm} && \frac{1}{2} \left\{ tr(V_j^{* \,-1} V_i^*) + \frac{n-t_i-t_0}{SSE_i}(\hat{y}_i^*-\hat{y}_j^*)^{\prime} V_j^{* \,-1} (\hat{y}_i^*-\hat{y}_j^*)
 + \log \left( \frac{|V_j^*|}{|V_i^*|}\right) -n^* \right\}.
\een
When it comes to computing the criterion OMD of formula (\ref{MDfinal}) in the paper, all terms $\log(|V_j^*|/|V_i^*|)$ disappear because the sum extends over all indexes $i \neq j$.
\newpage
\subsection*{Appendix B: posterior distribution of $(\beta_0, \beta_i, \sigma)$  under the reference and the hierarchical $g$-prior}
Consider the linear model $M_i$  represented by equation (\ref{linearmodelMi}) in the paper, and assume for simplicity that $\beta_0$ is a scalar ($t_0=1$).
We want to show that   the posterior distribution of $(\beta_0, \beta_i, \sigma)$  under  the hierarchical $g$-prior can be approximated with the corresponding distribution under the reference prior, at least when $n$ is moderately large.
 Consider first the posterior under the standard reference prior $p^N(\beta_0, \beta_i, \sigma \g  M_i) \propto 1/\sigma$. This is given by
\begin{eqnarray*}
\label{p^NbetaPosterior}
p^R(\beta_0, \beta_i, \sigma \g y,  M_i)=N(\beta_0 \g \bar{y}, \sigma^2/n)
N_{t_i}(\beta_i \g \hat{\beta}_i, \sigma^2(V_i^{\prime}V_i)^{-1})
IGa(\sigma^2 \g \frac{n-t_i-1}{2}, \frac{SSE_i}{2}).
\end{eqnarray*}
On the other hand, if the prior is the hierarchical $g$-prior, see equation (\ref{p^R}) in the paper, the posterior becomes
\begin{eqnarray*}
\label{p^RbetaPosterior}
\hspace{-1cm}&&p^R(\beta_0, \beta_i, \sigma \g y,  M_i)=N(\beta_0 \g \bar{y}, \sigma^2/n) \nonumber \\
\hspace{-1cm}&&\int N_{t_i}(\beta_i \g \frac{g}{g+1}\hat{\beta}_i, \frac{g}{g+1} \sigma^2(V_i^{\prime}V_i)^{-1})
IGa(\sigma^2 \g \frac{n-1}{2}, \frac{g}{2(g+1)}(SSE_i+\frac{1}{g}SSE_0))p^R(g \g M_i)dg.
\end{eqnarray*}
Since $p^R(g \g M_i)$  is positive only for $g> \frac{1+n}{t_i+t_0}-1$, it follows that as $n$ grows, so does $g$ in probability; in particular $\frac{g}{g+1} \stackrel{p} \rightarrow 1$ ($n \rightarrow \infty$), and the two posterior distributions  become  similar.
The above argument was  developed in a preliminary version of the article \citet{Baya:Berg:Fort:Garc:2012}, but is not present in the final version of the paper.

\newpage

\subsection*{Tables and Figures}

\renewcommand{\baselinestretch}{1}
\begin{table}[htbp] \footnotesize
\centering
\caption{{\protect\small \textit{Injection molding experiment.
Candidate follow-up runs}}}
\label{tab:Injection.Candidate.Runs}
\begin{tabular}{*{6}{c}}
\hline
Run in the $2^{4}$ full design  & A & C & E & H & Corresponding runs in the $2^{8-4}$ fractional design \\
\hline $1$ & $-$ & $-$ & $-$ & $-$ & 14,16 \\
$2$ & $-$ & $-$ & $+$ & $+$ & 1,3 \\
 $3$ & $-$ & $+$ & $-$ & $+$ & 5,7 \\
 $4$ & $-$ & $+$ & $+$ & $-$ & 10,12 \\
 $5$ & $+$ & $-$ & $-$ & $+$ & 2,4 \\
 $6$ & $+$ & $-$ & $+$ & $-$ & 13,15 \\
 $7$ & $+$ & $+$ & $-$ & $-$ & 9,11 \\
 $8$ & $+$ & $+$ & $+$ & $+$ & 6,8 \\
 $9$ & $-$ & $-$ & $-$ & $+$ &  \\
 $10$ & $-$ & $-$ & $+$ & $-$ &  \\
 $11$ & $-$ & $+$ & $-$ & $-$ &  \\
 $12$ & $-$ & $+$ & $+$ & $+$ &  \\
 $13$ & $+$ & $-$ & $-$ & $-$ &  \\
 $14$ & $+$ & $-$ & $+$ & $+$ & \\
 $15$ & $+$ & $+$ & $-$ & $+$ &  \\
 $16$ & $+$ & $+$ & $+$ & $-$ &  \\

\hline
\end{tabular}%
\end{table}




\renewcommand{\baselinestretch}{1}
\begin{table}[htbp] \footnotesize
\centering
\caption{{\protect\small \textit{Reactor experiment.
Full $2^{5}$ factorial design}}}
\label{tab:Reactor.Full.Fact.Design}%
\begin{tabular}{*{9}{c}}
\hline
Run   & & A & B & C & D & E & &  $y$ \\
\hline $1$  & & $-$ & $-$ & $-$ & $-$ & $-$ & & 61 \\
 $2$ & & $+$ & $-$ & $-$ & $-$ & $-$ & &53\\
 $3$ & & $-$ & $+$ & $-$ & $-$ & $-$ & & 63\\
 $4$ & & $+$ & $+$ & $-$ & $-$ & $-$ & & 61\\
 $5$ & & $-$ & $-$ & $+$ & $-$ & $-$ & & 53\\
 $6$ & & $+$ & $-$ & $+$ & $-$ & $-$ & & 56\\
 $7$ & & $-$ & $+$ & $+$ & $-$ & $-$ & & 54\\
 $8$ & & $+$ & $+$ & $+$ & $-$ & $-$ & & 61\\
 $9$ & & $-$ & $-$ & $-$ & $+$ & $-$ & & 69\\
 $10$ & & $+$ & $-$ & $-$ & $+$ & $-$ & & 61\\
 $11$ & & $-$ & $+$ & $-$ & $+$ & $-$ & & 94\\
 $12$ & & $+$ & $+$ & $-$ & $+$ & $-$ & & 93\\
 $13$ & & $-$ & $-$ & $+$ & $+$ & $-$ & & 66\\
 $14$ & & $+$ & $-$ & $+$ & $+$ & $-$ & & 60\\
 $15$ & & $-$ & $+$ & $+$ & $+$ & $-$ & & 95\\
 $16$ & & $+$ & $+$ & $+$ & $+$ & $-$ & & 98\\
 $17$ & & $-$ & $-$ & $-$ & $-$ & $+$ & & 56\\
 $18$ & & $+$ & $-$ & $-$ & $-$ & $+$ & & 63\\
 $19$ & & $-$ & $+$ & $-$ & $-$ & $+$ & & 70\\
 $20$ & & $+$ & $+$ & $-$ & $-$ & $+$ & & 65\\
 $21$ & & $-$ & $-$ & $+$ & $-$ & $+$ & & 59\\
 $22$ & & $+$ & $-$ & $+$ & $-$ & $+$ & & 55\\
 $23$ & & $-$ & $+$ & $+$ & $-$ & $+$ & & 67\\
 $24$ & & $+$ & $+$ & $+$ & $-$ & $+$ & & 65\\
 $25$ & & $-$ & $-$ & $-$ & $+$ & $+$ & & 44\\
 $26$ & & $+$ & $-$ & $-$ & $+$ & $+$ & & 45\\
 $27$ & & $-$ & $+$ & $-$ & $+$ & $+$ & & 78\\
 $28$ & & $+$ & $+$ & $-$ & $+$ & $+$ & & 77\\
 $29$ & &  $-$ & $-$ & $+$ & $+$ & $+$ & & 49\\
 $30$ & & $+$ & $-$ & $+$ & $+$ & $+$ & & 42\\
 $31$ & & $-$ & $+$ & $+$ & $+$ & $+$ & & 81\\
 $32$ & & $+$ & $+$ & $+$ & $+$ & $+$ & & 82\\
\hline
\end{tabular}%
\end{table}

\renewcommand{\baselinestretch}{1}
\begin{table}[htbp] \footnotesize
\centering
\caption{{\protect\small \textit{Reactor experiment.
Screening design}}}

\label{tab:Reactor.Screening.Design}%
\begin{tabular}{*{9}{c}}
\hline
Run in the full design & Run & A & B & C & D & E & & $y$ \\
\hline $2$ & $1$ & $+$ & $-$ & $-$ & $-$ & $-$ && 53 \\
$7$ & $2$ & $-$ & $+$ & $+$ & $-$ & $-$ & &54\\
 $12$ & $3$ & $+$ & $+$ & $-$ & $+$ & $-$ &&93\\
 $13$ & $4$ & $-$ & $-$ & $+$ & $+$ & $-$ & &66\\
 $19$ & $5$ & $-$ & $+$ & $-$ & $-$ & $+$ & &70\\
 $22$ & $6$ & $+$ & $-$ & $+$ & $-$ & $+$ & &55\\
 $25$ & $7$ & $-$ & $-$ & $-$ & $+$ & $+$ & &44\\
 $32$ & $8$ & $+$ & $+$ & $+$ & $+$ & $+$ & & 82\\
\hline
\end{tabular}%
\end{table}


\renewcommand{\baselinestretch}{1}
\begin{table}[htbp] \footnotesize
\centering
\caption{{\protect\small \textit{Reactor experiment.
Posterior probabilities of top five models and that factors are active (3FI)}}}
\label{tab:Reactor.Post.Prob.TopModelsAndActFac.3FI}
\begin{tabular}{*{6}{c}}
\hline
\hline
 & \multicolumn{2}{c}{Conventional Approach} & & \multicolumn{2}{c}{Objective Approach} \\
\cline{2-3}
\cline{5-6}
 Model & Factors   &  Posterior probability & & Factors & Posterior probability  \\
\hline $1$ & $null$  & $0.23$ & & $null$ & $0.46$\\
 $2$ & $B$  & $0.13$ & &$B$  & $0.12$ \\
 $3$ & $D$  & $0.07$ && $A,D$  & $0.07$\\
 $4$ & $A$  & $0.07$ && $B,D$  & $0.07$ \\
 $5$ & $A,B$  & $0.05$ && $A,B$  & $0.07$ \\
\hline
\hline
& \multicolumn{2}{c}{Conventional Approach} & & \multicolumn{2}{c}{Objective Approach} \\
\cline{2-3}
\cline{5-6}
 &  Factor & Posterior probability &  &  Factor & Posterior probability  \\
\cline{2-6}
& $A$ & $0.27$  && $A$ & $0.20$ \\
 & $B$ & $0.37$  & &$B$ & $0.31$ \\
 & $C$ & $0.17$   && $C$ &  $0.06$\\
 & $D$ & $0.29$  & &$D$ & $0.21$  \\
 & $E$ & $0.17$   && $E$ & $0.06$  \\
 \hline
\end{tabular}%
\end{table}

\renewcommand{\baselinestretch}{1}
\begin{table}[htbp] \footnotesize
\centering
\caption{{\protect\small \textit{Reactor experiment.
Top five follow-up designs (3FI)}}}
\label{tab:Reactor.Top.Five.Follow-up.3FI}
\begin{tabular}{*{7}{c}}
\hline
\hline
Model & & CMD & runs  & & OMD  & runs \\
\cline{1-1}
\cline{3-4}
\cline{6-7}
 $1$ && $0.6535$ & 4  10 11 28 && $1.5647$ & 4  10 11 28   \\
 $2$ && $0.6529$ & 4  10 11 12  && $1.5625$ &  4 26 27 28    \\
$3$ && $0.6502$ & 10 11 12 26  &&$1.5624$ &  20 26 27 28   \\
$4$ && $0.6501$ & 10  12 26 27  && $1.5623$ &  4 10 16 28   \\
 $5$ && $0.6499$ & 4 10 12 26  && $1.5610$ &  4 11 26 28   \\
\hline
\end{tabular}%
\end{table}


\renewcommand{\baselinestretch}{1}
\begin{table}[tbp] \footnotesize
\centering
\caption{{\protect\small \textit{Reactor experiment.
Posterior probabilities of top five models and that factors are active based on  the combined screening and follow-up designs designs (3FI)}}}
\label{tab:Reactor.Post.Prob.Combined.3FI}%
\begin{tabular}{*{6}{c}}
\hline
\hline
 & \multicolumn{2}{c}{Conventional Approach} & & \multicolumn{2}{c}{Objective Approach} \\
\cline{2-3}
\cline{5-6}
 Model & Factors   &  Posterior probability & & Factors & Posterior probability  \\
\hline $1$ & $B,D,E$ & $0.38$ & & $null$ & $0.27$\\
 $2$ & $B,D$  & $0.25$ && $B,D,E$  & $0.21$ \\
 $3$ & $null$  & $0.11$ && $B,D$  & $0.20$\\
 $4$ & $B$  & $0.11$ && $B$  & $0.10$\\
 $5$ & $B,C,D,E$  & $0.05$ && $D$  & $0.04$\\
\hline
\hline
& \multicolumn{2}{c}{Conventional Approach} & & \multicolumn{2}{c}{Objective Approach} \\
\cline{2-3}
\cline{5-6}
 &  Factor & Posterior probability &  &  Factor & Posterior probability  \\
\cline{2-6}
 & $A$ & $0.03$  && $A$ & $0.09$ \\
 & $B$ & $0.82$  && $B$ & $0.62$ \\
 & $C$ & $0.08$  & & $C$ &  $0.08$\\
 & $D$ & $0.74$  && $D$ & $0.53$  \\
 & $E$ & $0.45$  & & $E$ & $0.27$  \\
\hline
\end{tabular}%
\end{table}





\begin{figure}[H] 
\begin{center}
\includegraphics[width=6in]{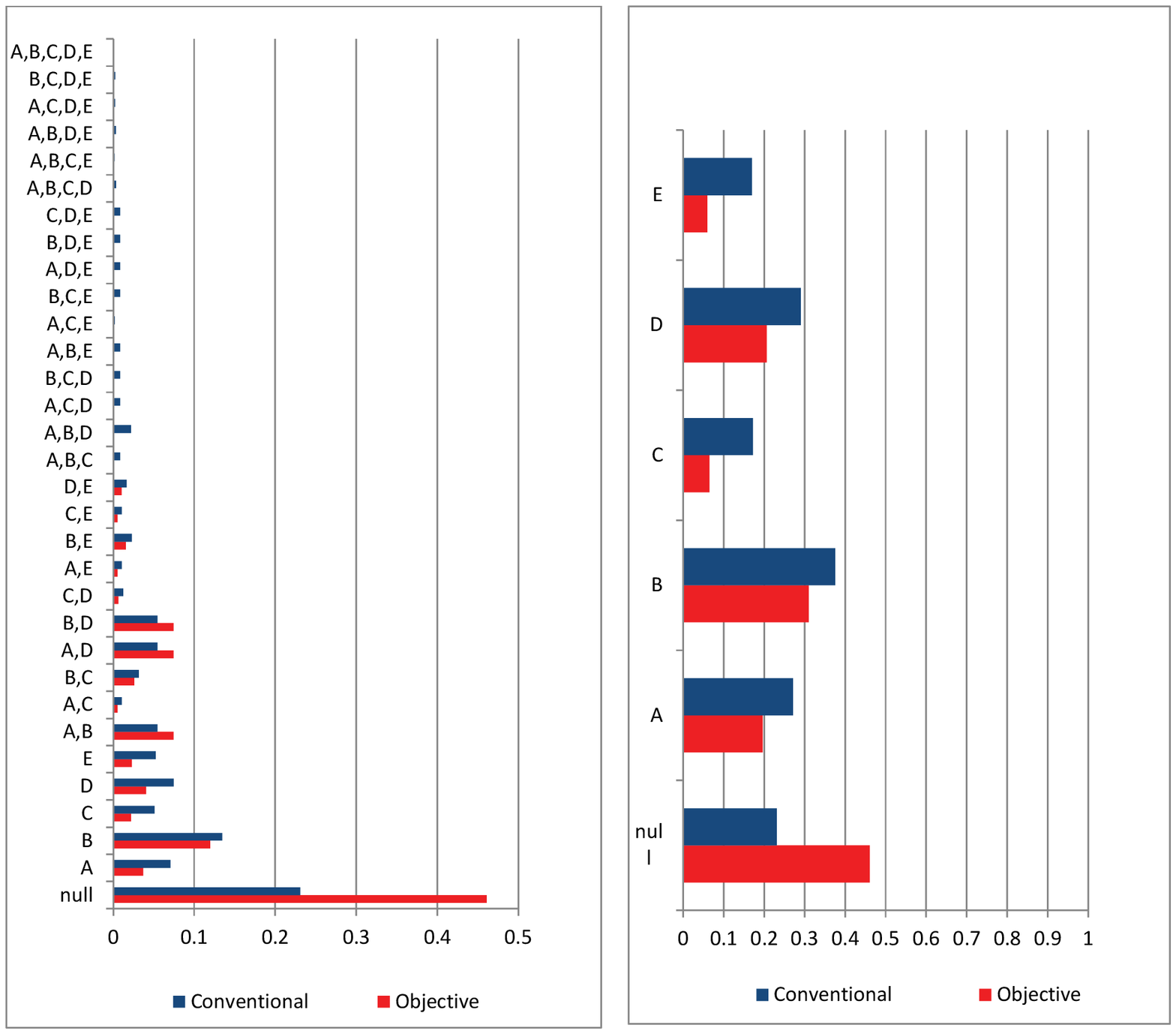}
\end{center}
\caption{\textit{Reactor experiment. Posterior  probabilities of models and that factors are active (3FI)}
\label{fig:Reactor.Post.Prob.Models.Factors.3FI} }
\end{figure}

\newpage

\begin{figure}[H]
\begin{center}
\includegraphics[width=6in]{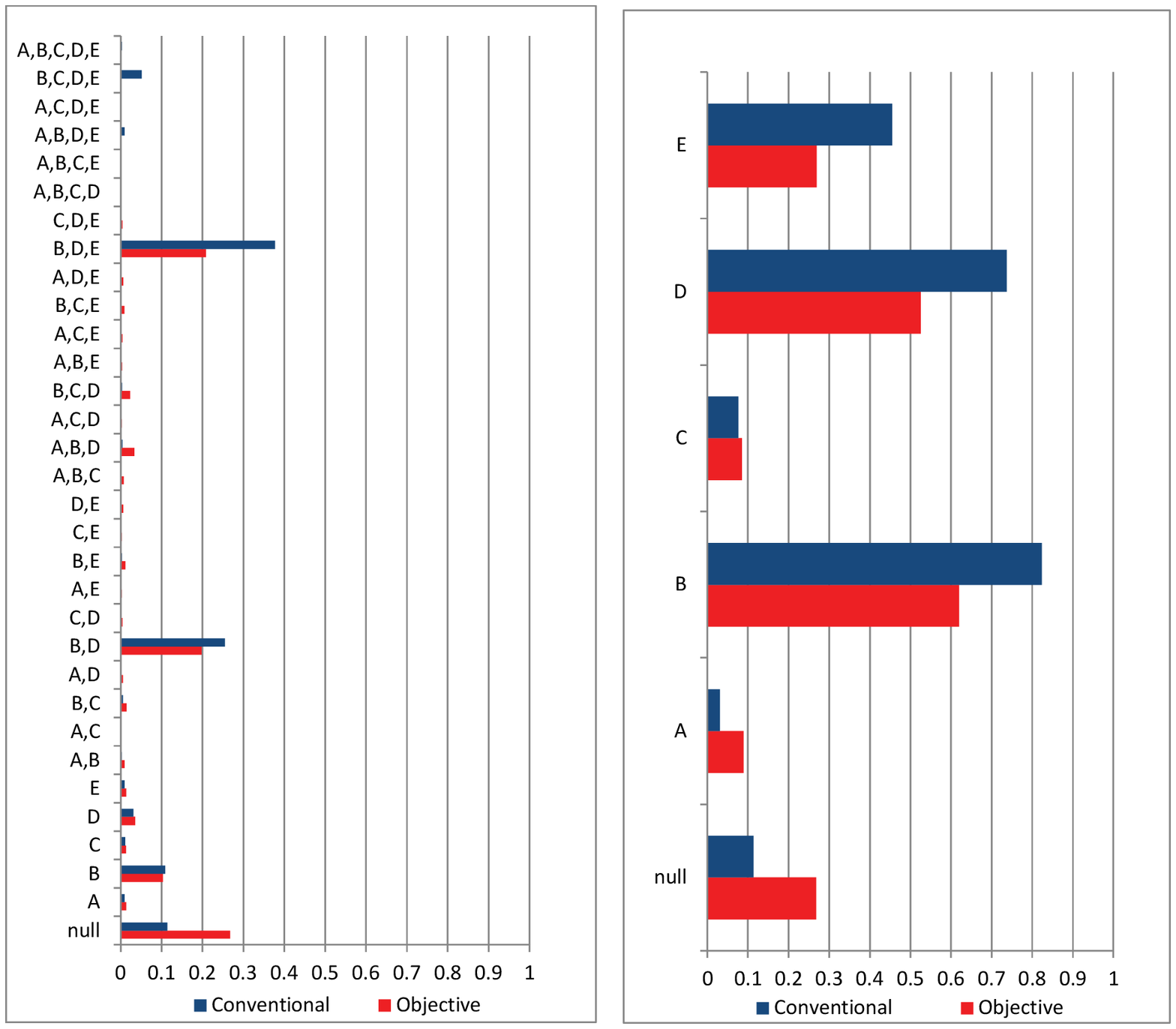}
\end{center}
\caption{\textit{Reactor experiment. Posterior  probabilities of models and that factors are active based on  the combined screening and follow-up designs (3FI)} }
\label{fig:Reactor.Post.Prob.Combined.3FI}
\end{figure}

\newpage

\begin{figure}[H]
\begin{center}
\includegraphics[scale=1]{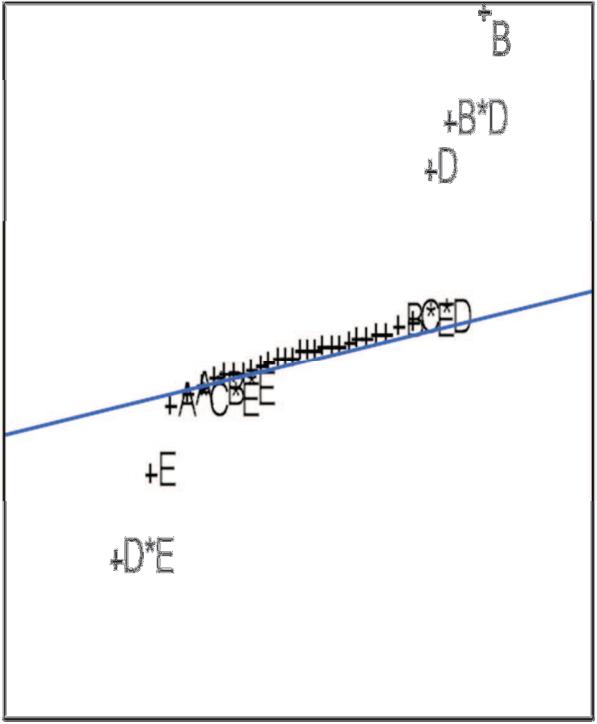}
\end{center}
\caption{\textit{Reactor experiment. Normal probability of  contrasts based on the \textit{complete} set of 32 runs}
\label{fig: Reactor.Normal.Prob.Contrasts} }
\end{figure}

\clearpage 
\bibliographystyle{biometrika}
\bibliography{BayesianDesign}

\begin{thebibliography}{26}
\expandafter\ifx\csname natexlab\endcsname\relax\def\natexlab#1{#1}\fi
\expandafter\ifx\csname url\endcsname\relax
  \def\url#1{\texttt{#1}}\fi
\expandafter\ifx\csname urlprefix\endcsname\relax\def\urlprefix{URL }\fi
\providecommand{\eprint}[2][]{\url{#2}}

\bibitem[{Abramowitz \& Stegun(1964)}]{Abra:Steg:1964}
\textsc{Abramowitz, M.} \& \textsc{Stegun, I.~A.} (1964).
\newblock \textit{Handbook of mathematical functions with formulas, graphs, and
  mathematical tables}, vol.~55 of \textit{National Bureau of Standards Applied
  Mathematics Series}.
\newblock U.S Government Printing Office, Washington, D.C.

\bibitem[{Barrios~Zamudio(2013)}]{Barr:2013}
\textsc{Barrios~Zamudio, E.} (2013).
\newblock Using the bsmd package for {B}ayesian screening and model
  discrimination.
\newblock \url{http://cran.r-project.org/web/packages/BsMD/vignettes/BsMD.pdf}.

\bibitem[{{Bayarri} et~al.(2012){Bayarri}, {Berger}, {Forte} \&
  {Garc{\'{\i}}a-Donato}}]{Baya:Berg:Fort:Garc:2012}
\textsc{{Bayarri}, M.~J.}, \textsc{{Berger}, J.~O.}, \textsc{{Forte}, A.} \&
  \textsc{{Garc{\'{\i}}a-Donato}, G.} (2012).
\newblock {Criteria for Bayesian model choice with application to variable
  selection}.
\newblock \textit{Annals of Statistics} 40 1550--1577.

\bibitem[{Bayarri \& Garcia-Donato(2007)}]{Baya:Garc:2007}
\textsc{Bayarri, M.~J.} \& \textsc{Garcia-Donato, G.} (2007).
\newblock Extending conventional priors for testing general hypotheses in
  linear models.
\newblock \textit{Biometrika} 94 135--152.

\bibitem[{Berger \& Pericchi(1996)}]{Berg:Peri:1996}
\textsc{Berger, J.~O.} \& \textsc{Pericchi, L.} (1996).
\newblock The intrinsic {B}ayes factor for model selection and prediction.
\newblock \textit{Journal of the American Statistical Association} 91 109--122.

\bibitem[{Berger \& Pericchi(2001)}]{Berg:Peri:2001}
\textsc{Berger, J.~O.} \& \textsc{Pericchi, L.~R.} (2001).
\newblock Objective {B}ayesian methods for model selection: introduction and
  comparison.
\newblock In \textit{Model selection}, vol.~38 of \textit{IMS Lecture Notes
  Monogr. Ser.} Beachwood, OH: Inst. Math. Statist., 135--207.

\bibitem[{Bingham \& Chipman(2007)}]{Bing:Chip:inco:2007}
\textsc{Bingham, D.~R.} \& \textsc{Chipman, H.~A.} (2007).
\newblock Incorporating {P}rior {I}nformation in {O}ptimal {D}esign for {M}odel
  {S}election.
\newblock \textit{Technometrics} 49 155--163.

\bibitem[{Box \& Hill(1967)}]{Box:Hill:1967}
\textsc{Box, G. E.~P.} \& \textsc{Hill, W.~J.} (1967).
\newblock Discrimination among mechanistic models.
\newblock \textit{Technometrics} 9 57--71.

\bibitem[{Box et~al.(1978)Box, Hunter \& Hunter}]{Box:Hunt:Hunt:1978}
\textsc{Box, G. E.~P.}, \textsc{Hunter, W.~G.} \& \textsc{Hunter, J.~S.}
  (1978).
\newblock \textit{Statistics for experimenters. An introduction to design, data
  analysis, and model building}.
\newblock John Wiley \& Sons, New York-Chichester-Brisbane.

\bibitem[{Casella \& Moreno(2006)}]{Case:More:2006}
\textsc{Casella, G.} \& \textsc{Moreno, E.} (2006).
\newblock Objective {B}ayesian variable selection.
\newblock \textit{Journal of the American Statistical Association} 101
  157--167.

\bibitem[{Castillo \& van~der Vaart(2012)}]{Cast:Vaar:2012}
\textsc{Castillo, I.} \& \textsc{van~der Vaart, A.} (2012).
\newblock Needles and straw in a haystack: posterior concentration for possibly
  sparse sequences.
\newblock \textit{Ann. Statist.} 40 2069--2101.

\bibitem[{Clyde et~al.(2011)Clyde, Ghosh \& Littman}]{Clyd:Ghos:Litt:2011}
\textsc{Clyde, M.~A.}, \textsc{Ghosh, J.} \& \textsc{Littman, M.~L.} (2011).
\newblock Bayesian adaptive sampling for variable selection and model
  averaging.
\newblock \textit{J. Comput. Graph. Statist.} 20 80--101.

\bibitem[{Han \& Chaloner(2004)}]{Han:Chal:2004}
\textsc{Han, C.} \& \textsc{Chaloner, K.} (2004).
\newblock Bayesian experimental design for nonlinear mixed-effects models with
  application to hiv dynamics.
\newblock \textit{Biometrics} 60 25--33.

\bibitem[{Jeffreys(1961)}]{Jeff:1961}
\textsc{Jeffreys, H.} (1961).
\newblock \textit{Theory of probability}.
\newblock Third edition. Clarendon Press, Oxford.

\bibitem[{Liang et~al.(2008)Liang, Paulo, Molina, Clyde \&
  Berger}]{Lian:EtAl:2008}
\textsc{Liang, F.}, \textsc{Paulo, R.}, \textsc{Molina, G.}, \textsc{Clyde,
  M.~A.} \& \textsc{Berger, J.~O.} (2008).
\newblock Mixtures of {$g$} priors for {B}ayesian variable selection.
\newblock \textit{J. Amer. Statist. Assoc.} 103 410--423.

\bibitem[{Maruyama \& George(2011)}]{Maru:Geor:2011}
\textsc{Maruyama, Y.} \& \textsc{George, E.~I.} (2011).
\newblock Fully {B}ayes factors with a generalized {$g$}-prior.
\newblock \textit{Ann. Statist.} 39 2740--2765.

\bibitem[{Meyer(1996)}]{Meye:1996}
\textsc{Meyer, D.} (1996).
\newblock mdopt: Fortran programs to generate md-optimal screening and
  follow-up designs, and analysis of data.
\newblock \url{http://lib.stat.cmu.edu/}.

\bibitem[{Meyer et~al.(1996)Meyer, Steinberg \& Box}]{Meye:Stei:Box:1996}
\textsc{Meyer, R.~D.}, \textsc{Steinberg, D.} \& \textsc{Box, G. E.~P.} (1996).
\newblock Follow-up designs to resolve confounding in fractional factorials.
\newblock \textit{Technometrics} 38 303--313.

\bibitem[{O'Hagan(1995)}]{Ohag:1995}
\textsc{O'Hagan, A.} (1995).
\newblock Fractional {B}ayes factors for model comparison.
\newblock \textit{J. Roy. Statist. Soc. Ser. B} 57 99--138.
\newblock With discussion and a reply by the author.

\bibitem[{O'Hagan \& Forster(2004)}]{Ohag:Fors:2004}
\textsc{O'Hagan, A.} \& \textsc{Forster, J.} (2004).
\newblock \textit{Kendall's Advanced Theory of Statistics, Vol. 2b: Bayesian
  Inference}.
\newblock Arnold, 2nd ed.

\bibitem[{Perez \& Berger(2002)}]{Pere:Berg:2002}
\textsc{Perez, J.~M.} \& \textsc{Berger, J.~O.} (2002).
\newblock Expected posterior prior distributions for model selection.
\newblock \textit{Biometrika} 89 491--512.

\bibitem[{Pericchi(2005)}]{Peri:2005}
\textsc{Pericchi, L.~R.} (2005).
\newblock Model selection and hypothesis testing based on objective
  probabilities and {B}ayes factors.
\newblock In D.~Dey \& C.~R. Rao, eds., \textit{Bayesian thinking: modeling and
  computation}, vol.~25 of \textit{Handbook of Statistics}.
  Elsevier/North-Holland, Amsterdam, 115--149.

\bibitem[{Scott \& Berger(2010)}]{Scot:Berg:2010}
\textsc{Scott, J.~G.} \& \textsc{Berger, J.~O.} (2010).
\newblock {B}ayes and empirical-{B}ayes multiplicity adjustment in the
  variable-selection problem.
\newblock \textit{Ann. Statist.} 38 2587--2619.

\bibitem[{Tsutakawa(1972)}]{Tsut:1972}
\textsc{Tsutakawa, R.~K.} (1972).
\newblock Design of experiment for bioassay.
\newblock \textit{Journal of the American Statistical Association} 67 584--590.

\bibitem[{Wolters \& Bingham(2011)}]{Wolt:Bing:2011}
\textsc{Wolters, M.~A.} \& \textsc{Bingham, D.~R.} (2011).
\newblock Simulated annealing model search for subset selection in screening
  experiments.
\newblock \textit{Technometrics} 53 225--237.

\bibitem[{Zellner \& Siow(1980)}]{Zell:Siow:1980}
\textsc{Zellner, A.} \& \textsc{Siow, A.} (1980).
\newblock Posterior odds ratios for selected regression hypotheses.
\newblock In J.~M. Bernardo, M.~H. DeGroot, D.~V. Lindley \& A.~F.~M. Smith,
  eds., \textit{Bayesian Statistics: Proceedings of the First International
  Meeting held in Valencia (Spain)}. University of Valencia, 585--603.

\end{thebibliography}

\end{document}